\documentclass[sigconf]{acmart}

\usepackage{microtype}
\usepackage{graphicx}
\usepackage{subcaption}
\usepackage{caption}

\setcopyright{rightsretained}
\copyrightyear{2025}
\acmYear{2025}
\acmDOI{10.5281/zenodo.14930029}
\acmISBN{}
\acmConference[3rd Workshop on Explainability in HRC]{3rd Workshop on Explainability in Human-Robot Collaboration at HRI '25}{March 3, 2025}{Melbourne, Australia}

\acmBooktitle{The 3rd Workshop on Explainability in Human-Robot Collaboration at HRI '25}

\begin{document}

\title{
    Enhancing Explainability with Multimodal Context Representations for Smarter Robots
}

\author{Anargh Viswanath}
\authornote{Both authors contributed equally to this research.}
\orcid{0009-0009-8342-1559}
\affiliation{%
  \institution{Bielefeld University}
  \city{Bielefeld}
  \country{Germany}
}
\email{anargh.viswanath@uni-bielefeld.de}

\author{Lokesh Veeramacheneni}
\authornotemark[1]
\orcid{0009-0003-6545-2083}
\affiliation{%
  \institution{University of Bonn}
  \city{Bonn}
  \country{Germany}
}
\email{lveerama@uni-bonn.de}

\author{Hendrik Buschmeier}
\orcid{0000-0002-9613-5713}
\affiliation{%
  \institution{Bielefeld University}
  \city{Bielefeld}
  \country{Germany}
}
\email{hbuschme@uni-bielefeld.de}

\begin{abstract}
    Artificial Intelligence (AI) has significantly advanced in recent years, driving innovation across various fields, especially in robotics. Even though robots can perform complex tasks with increasing autonomy, challenges remain in ensuring explainability and user-centered design for effective interaction. A key issue in Human-Robot Interaction (HRI) is enabling robots to effectively perceive and reason over multimodal inputs, such as audio and vision, to foster trust and seamless collaboration. In this paper, we propose a generalized and explainable multimodal framework for context representation, designed to improve the fusion of speech and vision modalities. We introduce a use case on assessing {\emph{‘Relevance’}} between verbal utterances from the user and visual scene perception of the robot. We present our methodology with a {\emph{Multimodal Joint Representation}} module and a {\emph{Temporal Alignment}} module, which can allow robots to evaluate relevance by temporally aligning multimodal inputs. Finally, we discuss how the proposed framework for context representation can help with various aspects of explainability in HRI.
\end{abstract}



\maketitle

\section{Introduction}

\begin{figure*}
    \centering
    \includegraphics[width=\linewidth]{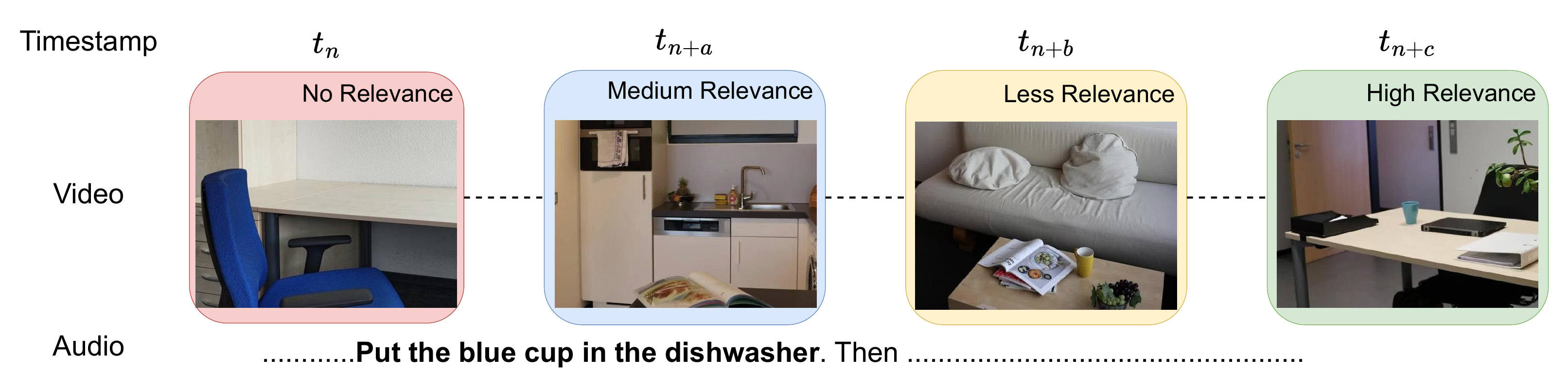}
    \caption{Illustrating \emph{‘Relevance’} metric for evaluating when and how well verbal utterances from users align with the robot’s visual perceptions. Here the verbal utterance \emph{`Put the blue cup in the dishwasher'} is of high relevance to the visual scene at time $t_{n+c}$, while the visual frame at time $t_{n}$  has no relevance due to the absence of relevant objects. Similarly, the third visual scene with \emph{`yellow cup'} at time $t_{n+b}$ is less relevant for the verbal utterance compared to the second visual scene with \emph{`dishwasher'}.}
    \Description{Illustration showing five boxes, each at subsequent points in time ($t_N$, $t_{n+a}$, $t_{n+b}$, $t_{n+c}$ and containing a different view into the scene, roughly time-aligned with an audio prompt. Each box shows a different degree of relevance according to the relevance metric.}
    \label{fig:intro_image}
\end{figure*}

Artificial Intelligence (AI) has seen rapid advancements in recent years, leading to its widespread adoption across various domains. Among these, the application of AI in robotics has brought transformative changes, enabling robots to perform complex tasks with greater autonomy and efficiency. However, despite these advancements, there are still significant challenges associated with them, particularly in terms of explainability and user-centered design. A fundamental issue in the field of Human-Robot Interaction (HRI) is enabling robots to perceive and understand their environment through multimodal inputs, such as audio and visual data. To interact effectively with humans, robots must be able to integrate and reason over multimodal information, ensuring that speech (verbal commands from the user) and vision (robot perception) are processed in a coherent and context-aware manner. Combining these inputs requires a structured approach to avoid information loss and create reasoning capabilities for fostering trust and long-term collaboration. This becomes particularly crucial in human-centered spaces, where robots are expected to comprehend routines, execute commands seamlessly, and exhibit reasoning capabilities when encountering uncertainties.

In this paper, we propose a generalized and explainable multimodal framework for context representation, designed to improve the fusion of speech and vision modalities in HRI-based applications. To demonstrate its effectiveness, we introduce a use case focused on assessing {\emph{‘Relevance’}}, a key metric that evaluates when and how well verbal utterances from users align with the robot’s visual perception. To illustrate our use case, consider a service robot deployed in a household setting for tidying and rearranging things. For completing these tasks, the robot can navigate autonomously and is equipped with sensors for continuously perceiving its environment within its field of view. Additionally, it can also receive verbal commands or utterances from humans through audio sensors. Suppose that while the robot is carrying out its routine cleaning, it receives a verbal command from the user {\emph{`Put the blue cup in the dishwasher'}} as shown in Fig.~\ref{fig:intro_image}. The user probably assumes that the robot will complete it as an additional task. However, in the current field of view as well as the previous visual scenes, it has encountered a \emph{`dishwasher'} but not a \emph{`blue cup'}. This necessitates the need for temporal alignment between the multimodal inputs to assess the relevance. In this paper, we propose a novel methodology consisting of a {\emph{Multimodal Joint Representation}} module along with a {\emph{Temporal Alignment}} module, which quantifies the relevance factor by temporally aligning multimodal representations. This will allow the robot to plan actions strategically and enhance its ability to bridge the gap between language and vision.

Lastly, we explore how the proposed multimodal framework can contribute to explainability through two key aspects, namely {\emph{Useful Representation}} and {\emph{Meaningful Abstraction}}. These factors play a critical role in improving transparency, interpretability, and user experience in HRI. This work is ongoing, and in this paper, we outline our methodology for multimodal context representation, with a particular focus on the \emph{‘Relevance’} problem.

\section{Combining Vision and Speech}

Humans possess an extraordinary ability to effortlessly ground verbal expressions in the physical world. We are seamlessly linking language to real-world objects. We intuitively merge speech and vision modalities, enabling us to interpret verbal utterances within the context of visual scenes and vice versa \cite{Mondada2016}. This innate ability supports social interaction along with collaborative tasks between humans \cite{mondada2014temporal}. However, achieving similar capabilities in robots remains a significant challenge in the field of HRI \cite{Allgeuer2024}.

Developing intelligent robots capable of long-term, meaningful interaction with humans, especially robots designed for collaborative tasks (e.g., service robots), requires overcoming this gap. In human-centred spaces, such as homes or healthcare settings, robots need to interact in a way that feels natural and sociable. This is essential for long-term and effective collaboration to build trust. For this, robots must possess certain capabilities, such as context awareness, multimodal knowledge representation, resolving references to objects in the physical world (grounding), and modality fusion for coherent reasoning, among others \cite{Allgeuer2024}.

\subsection{Traditional Representations}

Traditionally, robots have relied on representations such as spatial representations (e.g. metric maps, topological maps, semantic maps, etc.) \cite{Landsiedel2017}, hierarchical structures (e.g. associated with behavior trees, state machines, etc.) \cite{Ghzouli2023}, and graphical models (e.g. knowledge graphs, scene graphs, etc.) \cite{ConvAI-graphs-2024} to represent world information \cite{Paulius2019}. These structures help robots plan and perform physical tasks by breaking down complex actions into simpler, manageable steps. 

These approaches have proven effective for tasks like navigation and object manipulation, offering advantages in logical inferencing, failure-detection, and error diagnosis. However, these representations suffer with significant limitations. Firstly, these representations are often heavily vision-reliant, making it difficult to incorporate rich multimodal data, such as audio-video, into their representation space. This results in separated/disconnected processing of speech and vision, leading to information loss and inadequate reasoning capabilities for tasks that demand social interaction. Secondly, they might struggle in out-of-distribution scenarios, where the addition of new elements, such as objects or relationships, increases complexity and reduces performance. They may prove to be inflexible to domain adaptation and require extensive remodeling for new environments or applications.

\subsection{Multimodal Representations}

In recent years, Machine Learning (ML) and Deep Learning (DL) have ushered in a new era for advanced robotics. The developments in these fields have empowered robots with superior efficiency and adaptive capabilities to perform complex tasks in dynamic environments \cite{Soori2023}. These fields have significantly helped improve the interaction capabilities of robots for collaborative tasks \cite{Semeraro2023}. 

Unlike traditional methods, Deep Learning (DL) architectures generate universal representations that are not dependent on a single modality. Given sufficient and diverse training data, these networks learn a generalized representation space. Moreover, with the techniques like few-shot learning \cite{song23fwl}, the learned representation space can be quickly adapted to out-of-distribution scenarios. Additionally, recent works also strongly advocate for DL based world models to learn robust joint representations \cite{lecun2022path, reed2022a}.

Although effective, DL suffers from two notable problems. The first is the large-scale dependency of Deep Neural Networks (DNNs) on data \cite{raghu21dovision}. We believe that this problem is somewhat contained, with large amounts of data on the Internet along with datasets being made open-source for utilization. However, we must emphasize that it is necessary to consider the ethical aspects while utilizing them for training. The second and most significant challenge is the black-box nature of the DNN. These networks often consist of a large number of layers and complex architectures, which inherently lack explainability. Existing explainable methods made substantial efforts in providing insights in DNN decision \cite{sundararajan17ig, selvaraju20gradcam, tenney20lit}. However, these methods are limited to individual domains of Computer Vision and Natural Language Processing. 

Additionally, we found a critical gap in the literature concerning the explainability in multimodal, specifically combining speech and vision representations. In a recent paper, \citep{li2024explainable} propose a DNN to learn a joint embedding multimodal representation space where the semantic structure of the data is encoded. Such a representation space would benefit from a mapping between modalities, further facilitating better alignment between modalities. Moreover, we also see that the existing explainability methods in DL are largely standalone as they rely on gradients, and do not explicitly capture the intrinsic reasoning behind the decisions. 

Given these challenges, we argue that there is a substantial need to integrate explainability as an integral part of DNN based representations for their effective adoption in the field of HRI. This paper makes an effort toward this direction and proposes a framework for generalized and explainable multimodal (speech and vision) representation within robots. Such representations can help robots exhibit intelligent behavior with multimodal context-awareness. Additionally, they can enhance explainability across multiple levels of abstraction and facilitate the integration of traditional symbolic methods with modern DNN representations.

\section{Use Case -- Finding Relevance}

To provide more clarity, we illustrate the benefits of the proposed multimodal representation for robots through a practical use-case of understanding `\emph{Relevance}'. We define relevance as the key parameter that measures the extent and timing of alignment between a verbal utterance and a visual scene, or vice versa. This can bridge the gap between language and vision modalities to enhance the robot's ability to interpret and respond effectively.

\subsection{Methodology}

\begin{figure}
    \centering
    \includegraphics[width=\linewidth]{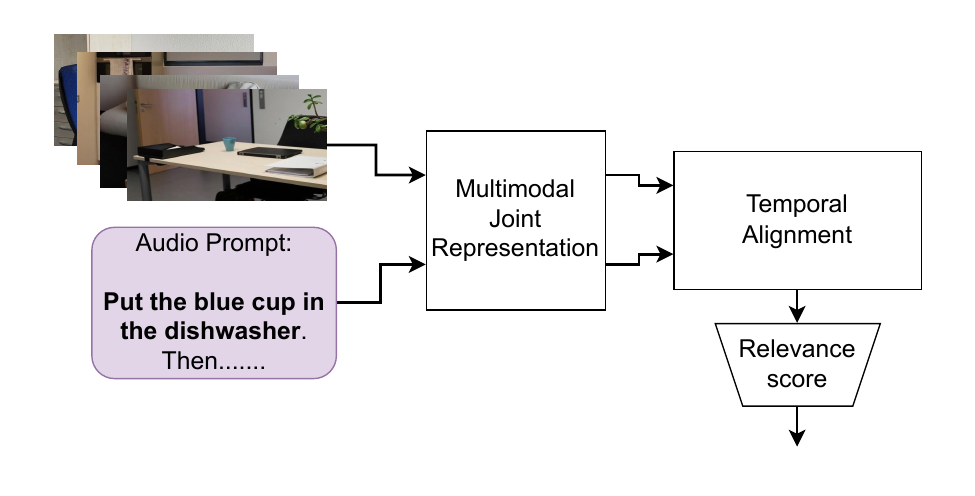}
    \caption{Schematic illustration of our proposed methodology with the video and speech inputs, where we first learn a joint representation space followed by temporally aligning the video-audio sequences which results in mapping between modalities and a quantified relevance metric for evaluation}
    \label{fig:methodology_image}
    \Description{Illustration of the proposed methodology. Video input (illustrated via multiple frames of a scene) and audio prompt (illustrated with the text "put the blue cup in the dishwasher. Then …" are fed into a box labeled Multimodal Joint Representation. This is fed forward into a box labeled Temporal Alignment, which is forwarded to a trapezoid box labeled Relevance score}
\end{figure}

For understanding the relevance factor, we propose a novel methodology with two modules, namely the \emph{Multimodal Joint Representation} and \emph{Temporal Alignment} module. Fig.~\ref{fig:methodology_image} demonstrates the proposed methodology. This subsection discusses individual modules in detail while providing reasons behind design decisions.

\subsubsection{Multimodal Joint Representation}

Typically, the visual and audio representation spaces are disjoint. In other words, mapping between these spaces is not linear. Prior works have utilized large-scale DNNs to learn this mapping between audio-video pairs by learning a joint representation space \cite{gurram2023lava, rouditchenko2021cascaded, rouditchenko21avlnet, hu2022mix, guzhov2022audioclip, Bain21frozenintime}. The DNNs in these works were trained in an unsupervised/self-supervised fashion, primarily to reduce annotations costs. A huge advantage of this joint representation approach is that it would inherently include the semantic meaning of objects in the representation space. Such a representation space can also be generalizable and be fine-tuned for a new domain.

Given the advantages provided by the joint representation space, we train a DNN to learn the combined representations of speech and visual domains. For example, in Fig.~\ref{fig:intro_image}, the \emph{`blue cup'} visual embedding and the corresponding \emph{`blue cup'} speech embeddings are supposedly similar. As an embedding is always in a lower dimension when compared to the original input space, computation is saved for further processing. Although effective, by design such a representation space does not cater for temporal alignment.

\subsubsection{Temporal Alignment}

To this end, we introduce a \emph{Temporal Alignment} module to temporally align joint-multimodal representation embeddings. For example, the verbal utterance \emph{`blue cup'} happens at around time $t_{n+a}$, whereas the visual appearance occurs at time $t_{n+c}$ (e.g. after the robot moves its camera). Given the embeddings for speech and visual, this module timely aligns these embeddings. Similar to the previous module and in accordance with previous works  \cite{lin2023univtg, dwibedi2019temporal, chowdhury2024meerkat, boggust2019grounding}, we train this \emph{Temporal Alignment} module in a self-supervised manner, compensating for the lack of annotated data. The output of this module is supposedly a temporally aligned audio-visual space. This creates a mapping between objects in verbal utterances and visual scenes. We further use this mapping to extract the relevance score between the audio-visual objects.

\subsection{Explainability}

Earlier, we described the utilization of the proposed framework of multimodal representation for quantifying relevance by assessing the alignment for inter-modality sequences. We believe this will be beneficial to significantly improve user experience in human-centered applications demanding high levels of context awareness. Additionally, we hypothesize that the proposed DL framework for multimodal context representation can help with various aspects of explainability.

\subsubsection{Representation}

A key advantage of the framework would be robust and explainable representations obtained from the multimodal embedding space. The representations obtained would be similar to the world representations mentioned by \cite{lecun2022path}. This embedding space enables a structured representation where audio and video data coherently coexist, allowing for a holistic and interpretable mapping of multimodal inputs. 
Offering a clear representation of how different modalities interact, the framework makes it easier to interpret the decisions made at the end of the pipeline. By modeling a coherent multimodal space, we ensure learned representations remain transparent yet adaptable for tasks like grounding, intent recognition, and behavior prediction.

The evaluation of the representations can be performed using metrics for alignment mentioned in prior studies such as Accuracy \cite{Wang_2020_WACV}, AV-Align \cite{yariv2024diverse}, Recall, and ROC-AUC \cite{han2022temporal}. In addition, qualitative user evaluations would also be beneficial. This can be carried out by either providing randomly selected verbal utterances and visual sequences with relevance scores for assessment or by deploying the modules on the robot to assess the interaction.

\subsubsection{Abstraction}

Another crucial aspect of explainability would be to create meaningful abstractions at multiple levels tailored to the specific applications. The framework can enhance interpretability for various stakeholders by structuring information at different levels of abstraction. 

\paragraph{Robot}
In the case of robots, abstraction will allow the robot to operate using high-level representations that encapsulate essential patterns and relationships instead of solely relying on raw data. They will help reduce latency while displaying context-appropriate behavior. For instance, by assessing the relevance factor (refer Fig.~\ref{fig:intro_image}), the robot can plan when to act and more importantly how to act. It can try to get additional information either by navigating and/or waiting for verbal commands if there is a low relevance score. Thus, the robot can rely on simple logical inferencing with the representations to make explainable decisions.

\paragraph{User}
When we consider the user, meaningful abstraction can help in bridging the gap between the robot's behavior/responses and its interpretation. With higher levels of abstraction, additional insights can be provided by the robot for query clarifications, context summarization for complex sequences, and adaptive response generation for effective HRI. This will help foster trust in robots as users gain better insights about their capabilities. To give an example with the proposed use case, if the robot receives a verbal command from the user for placing the \emph{`blue cup'} in the \emph{`dishwasher'} and it encounters a \emph{`yellow cup'}. Using the relevance as metric, the robot can postpone its action to obtain more information and/or also ask for clarification from the user when it finds that the verbal utterance has little/no relevance to the visual scene (see Fig.~\ref{fig:intro_image}).

\paragraph{Developer}
From the developer's perspective, the proposed framework can enable clearer debugging, model optimization, and refinement through structured abstraction. Rather than dealing with opaque and low-level representations, developers can analyze performance at various levels making it easier to trace errors, adjust models, and fine-tune system behavior. For instance, by utilizing the relevance factor, the performance of the temporal alignment module and multimodal joint representation can be assessed (refer Fig.~\ref{fig:methodology_image}). Additionally, it can be beneficial in deciphering the reasoning if there is an unexpected behavior showcased by the robot deviating from the expected course of action. It can also help in observing the real-time decision-making process. Overall, the deeper levels of abstraction will help with explainability and assist in improving user experiences.

Meaningful abstractions can only be evaluated through qualitative assessments conducted by users and developers.

\section{Conclusion}

As DL-based techniques continue to progress, ensuring explainability and effective multimodal fusion remains a key challenge, particularly in the field of HRI. For robots to interact effectively with humans, they must be capable of integrating and reasoning over multimodal information, ensuring that speech (verbal commands from the user) and vision (robot perception) are processed coherently. Here, we propose a generalized and explainable multimodal (speech and vision) framework for context representation. 

To demonstrate the effectiveness of the proposed framework, we introduced a use case focused on ‘\emph{Relevance}’ -- defined as when and how well verbal utterances from users align with the robot’s visual perception. We present our methodology for assessing relevance with a \emph{Multimodal Joint Representation} module and a \emph{Temporal Alignment} module, which can allow robots to evaluate relevance by temporally aligning multimodal inputs. We discuss how the proposed multimodal framework contributes to explainability through two key aspects, namely useful \emph{Representation} and meaningful \emph{Abstraction}. Firstly, our multimodal embedding space provides a structured and interpretable representation where speech and visual data coherently coexist. Secondly, abstraction at multiple levels improves explainability for different stakeholders, such as the robot, user, and developer. 

This paper presents our ongoing research on multimodal context representation, with a focus on tackling the ‘\emph{Relevance}’ problem and detailing our proposed methodology. Our paper contributes to the existing literature by proposing a novel framework along with a meaningful use case to encourage the adoption of explainable multimodal AI in the field of HRI for future research.

\section*{Ethical Impact Assessment}

The objective of this paper is to advance the field of HRI using multimodal ML methods. Even though there can be potential societal consequences based on the ideas mentioned, we have not identified any that are specific to this paper.

\begin{acks}
    This research is funded/supported by 
    (i) the \grantsponsor{mwike}{Ministry of Economic Affairs, Industry, Climate Action and Energy of the State of North Rhine-Westphalia, Germany (MWIKE)}{https://www.wirtschaft.nrw/} in the it's OWL project `Hybrid Living', supervised by Project Management Jülich (PtJ),
    (ii) the \grantsponsor{bmbf}{German Federal Ministry of Education and Research (BMBF)}{https://www.bmbf.de/}:
    \grantnum[]{bmbf}{01IS22094A WEST-AI}, and 
    (iii) the \grantsponsor{dfg}{German Research Foundation (DFG)}{https://www.dfg.de}:
    \grantnum{dfg}{TRR 318/1 2021 – 438445824}. Responsibility for the content of this publication lies with the authors.
\end{acks}

\bibliographystyle{ACM-Reference-Format}
\balance
\bibliography{bibliography}

\end{document}